\def\figsize{9.5cm}

\def\ZoomFig{4}

\def\rn{}
\def\nn#1 #2{#2. #1}				
\def\nnn#1 #2 #3{#2. #3. #1}			
\def\nnnn#1 #2 #3 #4{#2. #3. #4 #1}		
\def\nnnnn#1 #2 #3 #4 #5{#2. #3. #4 #5. #1}	
\def\dualand{ and\hbox{ }}				
\def\multiand{, and\hbox{ }}				
\def\rf#1;#2;#3;#4;#5 {{\frenchspacing\par\rn#1, #3 {\bf #4}, #5 (#2). \par}}
\def\rg#1;#2;#3;#4;#5;#6 {{\frenchspacing\par\rn#1, #3 {\bf #4}, #5 (#2). \par}}
\def\rfbook#1;#2;#3;#4;#5 {{\frenchspacing\par\rn#1, {\it #3} (#5, #4, #2).\par}}
\def\rfprep#1;#2;#3 {{\par\frenchspacing\rn#1, #3 (#2).\par}}
\def\rfproc#1;#2;#3;#4;#5;#6 {{\frenchspacing\par\rn#1 #2, in {\it #3}, ed. #4 (#5: #6)\par}}
\def\rfprocp#1;#2;#3;#4;#5;#6;#7 {{\frenchspacing\par\rn#1 #2, in {\it #3}, ed. #4 (#5: #6), p#7\par}}

\def\rg#1;#2;#3;#4;#5;#6 {\par\rn#1 #2, {\it #3}, {\bf #4}, #5 (``#6'') \par}
\def\rf#1;#2;#3;#4;#5 {\par\rn#1, {\it #3}, {\bf #4}, #5 (#2)\par}
\def\rfbook#1;#2;#3;#4;#5 {{\frenchspacing\par\rn#1, {\it #3} (#4: #5, #2)\par}}
\def\rfproc#1;#2;#3;#4;#5;#6 {{\frenchspacing\par\rn#1 #2, in {\it #3}, ed. #4 (#5: #6)\par}}
\def\rfprocp#1;#2;#3;#4;#5;#6;#7 {{\frenchspacing\par\rn#1 #2, in {\it #3}, ed. #4 (#5: #6), p#7\par}}
\def\rfprep#1;#2;#3  {{\par\rn#1, #3, #2\par}}
\def\rfprepp#1;#2;#3 {{\par\rn#1 #2, #3\par}}

\def\etal{{\frenchspacing\it et al.}}

\def\eg{{\frenchspacing\it e.g.}}
\def\etc{{\frenchspacing\it etc.}}

\def\beq#1{\begin{equation}\label{#1}}
\def\eeq{\end{equation}}
\def\beqa#1{\begin{eqnarray}\label{#1}}
\def\eeqa{\end{eqnarray}}

\def\fig#1{Figure~\ref{#1}}

\def\spose#1{\hbox to 0pt{#1\hss}}
\def\simlt{\mathrel{\spose{\lower 3pt\hbox{$\mathchar"218$}}
     \raise 2.0pt\hbox{$\mathchar"13C$}}}
\def\simgt{\mathrel{\spose{\lower 3pt\hbox{$\mathchar"218$}}
     \raise 2.0pt\hbox{$\mathchar"13E$}}}
\def\simpropto{\mathrel{\spose{\lower 3pt\hbox{$\mathchar"218$}}
     \raise 2.0pt\hbox{$\propto$}}}

\def\ed{\end{document}}

\def\ns{{n_s}}
\def\nt{{n_t}}

\def\rhol{\rho_\Lambda}
\def\thetaqcd{\theta_{\rm qcd}}

\def\beq#1{\begin{equation}\label{#1}}
\def\eeq{\end{equation}}
\def\beqa#1{\begin{eqnarray}\label{#1}}
\def\eeqa{\end{eqnarray}}

\def\Otot{\Omega_{\rm tot}}

\def\ns{{n_s}}
\def\nt{{n_t}}

\def\dH{Q}
\def\dHt{Q_t}

\def\ng{n_\gamma}

\def\rhob{\rho_{\rm b}}
\def\rhoc{\rho_{\rm c}}

\def\rhon{\rho_\nu}
\def\rhol{\rho_\Lambda}

\def\xib{\xi_{\rm b}}
\def\xic{\xi_{\rm c}}

\def\xin{\xi_\nu}

\def\tento#1{\times 10^{#1}}

\documentclass[twocolumn,amsmath,nofootinbib]{revtex4} 
\usepackage{url}
\begin{document}
\input{epsf.sty}

\def\mit{1}
\def\penn{2}

\def\affilmrk#1{$^{#1}$}
\def\affilmk#1#2{$^{#1}$#2;}

\title{HOW DID IT ALL BEGIN?}

\author{
Max Tegmark\affilmrk{\mit}
}
\address{
\affilmk{\mit}{Dept.~of Physics, Massachusetts Institute of Technology, 
Cambridge, MA 02139}
}

\date{June 20, 2005. Essay for Young Researchers Competition in honor of Charles Townes.}

\begin{abstract}
How did it all begin? Although this question has undoubtedly lingered for as long as humans 
have walked the Earth, the answer still eludes us.
Yet since my grandparents were born, scientists have been able to refine this question to a degree I find truly remarkable.
In this brief essay, I describe some of my own past and ongoing work on this topic, 
centering on cosmological inflation. I focus on 
(1) observationally testing whether this picture is correct and
(2) working out implications for the nature of physical reality
(\eg, the global structure of spacetime, dark energy and our cosmic future, parallel universes and 
fundamental versus environmental physical laws).
(2) clearly requires (1) to determine whether to believe the conclusions.
I argue that (1) also requires (2), since it affects the probability calculations for inflation's 
observational predictions.
\end{abstract}


  
\maketitle

\setcounter{footnote}{0}

\subsection{The question refined, I}

How did it all begin? Although this physics question has undoubtedly lingered for as long as humans 
have walked the Earth, the answer still eludes us.
Yet since my grandparents were born, scientists have been able to refine this question to a degree I find truly remarkable.

\begin{figure} 
\centerline{\epsfxsize=8.5cm\epsffile{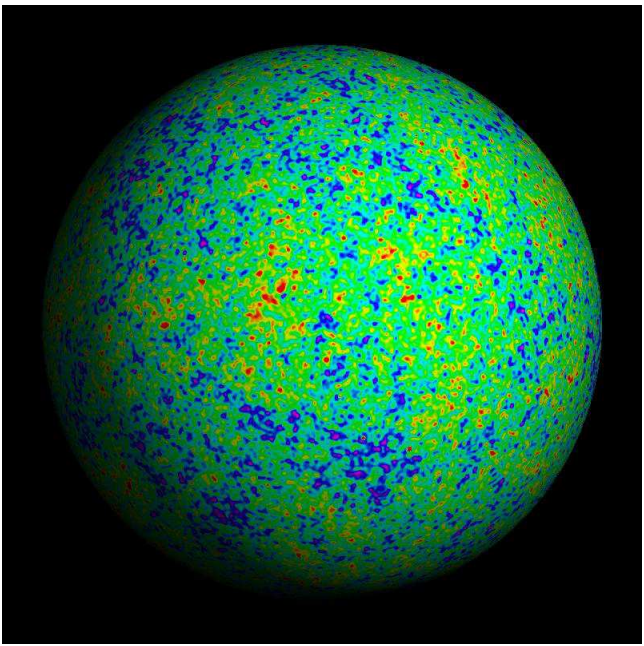}}
\caption[1]{\label{SphereFig}\footnotesize%
Our entire observable universe is inside this
sphere. Space continues outside
the sphere, but an opaque glowing wall of
hydrogen plasma hides it from our view. This
is our best image so far of what this plasma
sphere looks like, from \cite{mapforegs}, after cleaning out Galactic
radio noise from the WMAP satellite
observations. It has taken this light from the sphere's surface (known as the Cosmic Microwave Background radiation, CMB)
over 13 billion years to reach us at the center.
}
\end{figure}

First, the notion of ``it all'' was dramatically expanded by Edwin Hubble's 1925 discovery 
that the contemporary ``universe'' of nearby stars was merely part of one galaxy among countless others \cite{Hubble25}.
Today, most astronomers casually use the word ``universe'' to denote the spherical volume shown in \fig{SphereFig},
the cosmic event horizon containing about $10^{78}$ atoms and everything else we can in principle observe.

\begin{figure} 
\centerline{\epsfxsize=\figsize\epsffile{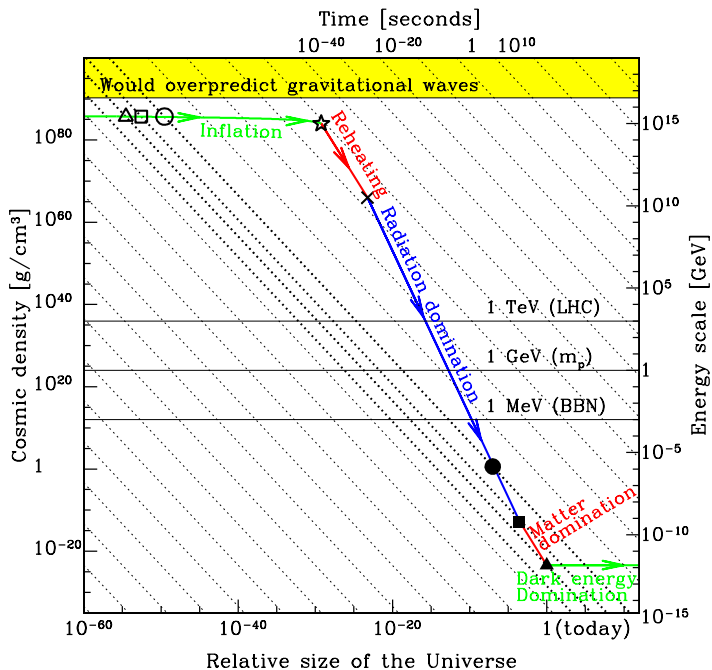}}
\caption[1]{\label{arhoFig}\footnotesize%
The cosmic mean density (solid curve) is diluted as the Universe expands.
Inflation is a period when there is almost no dilution, causing the expansion to accelerate,
and corresponds to the curve decreasing slower than the dotted diagonal lines of slope $-2$.
The two triangles lie on the same diagonal, which means that quantum fluctuations  
generated during inflation at the open triangle have been stretched into the 
horizon-scale fluctuations that we observe today at the filled triangle in the CMB (\fig{SphereFig}).
Detecting inflationary gravitational waves with CMB polarization would 
directly measure the shape of this cosmic density curve in the upper left corner of the plot,
just as the dark energy experiments directly measure the same curve 
in the lower right corner.
}
\end{figure}

Second, it has become clear that this universe is not static, 
but dynamic and evolving.
Spectacular recent measurements enabled by detector, computer and space technology have brought us a consistent
quantitative picture of how our universe expanded and evolved from a hot, fiery event known as the Big Bang some 14 billion years ago.
Our universe has expanded ever since the Big Bang, 
and this continuous stretching of space has both diluted and cooled the particles permeating it
(\fig{arhoFig}). As everything cooled, particles combined into progressively
more complex structures. Quarks combined to form protons and neutrons. Later, when the cosmic temperature was comparable to the
core of a star, fusion reactions combined neutrons and some of the protons into light elements like helium, deuterium and lithium.
About 400,000 years after the Big Bang, the leftover protons combined with electrons to form electrically neutral hydrogen
atoms, making the cosmos essentially transparent to light.
Up until this point, matter was extremely uniform, with only tiny $10^{-5}$-level density variations from place to place,
but gravitational attraction gradually clumped atoms together into galaxies, stars and planets, allowing 
atoms to form complex structures like molecules, cells, people and societies.

By the time I was born, the question ``How did it all begin?'' had thus been refined to inquiring about what happened when our 
universe was less than a second old. This included some particularly disturbing sub-questions.
For instance, why is space so big, so old and so flat, when generic initial conditions predict the
curvature to grow over time and the density to fast approach either zero or infinity 
(the ``flatness problem'')? 
What mechanism generated the $10^{-5}$ level ``seed'' fluctuations 
(visible as cosmic microwave background fluctuations in \fig{SphereFig})
out of which all cosmic structure grew,
and what conspiracy caused these fluctuations to have nearly identical amplitude in regions
of space that had never been in causal contact (the ``horizon problem'')?

\subsection{The question refined, II}

\begin{figure}[pbt]
\centerline{{\vbox{\epsfxsize=9.0cm\epsfbox{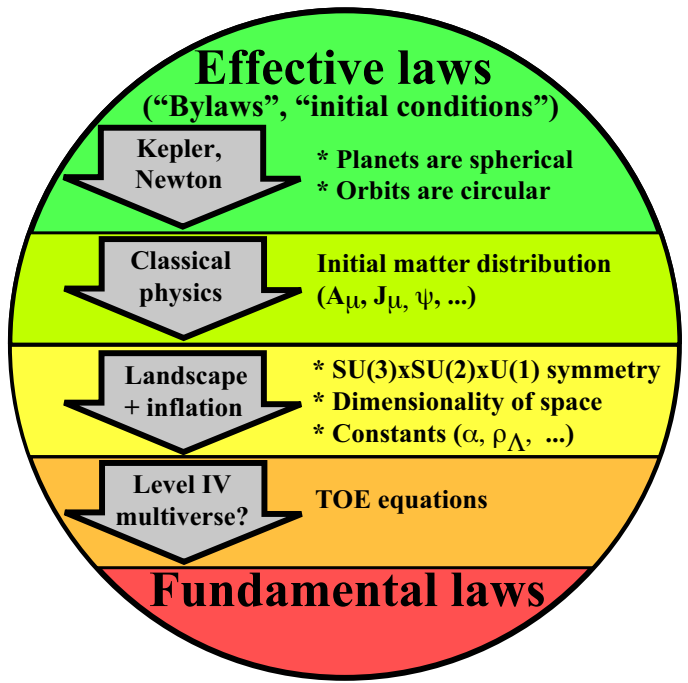}}}}

\caption{\label{BoundaryFig}
The shifting boundary (horizontal lines) 
between fundamental laws and environmental laws/effective laws/initial conditions.
Whereas Ptolemy and others had argued that the circularity of
planetary orbits was a fundamental law of nature, Kepler and Newton reclassified this as
an initial condition, showing that the fundamental laws also 
allowed highly non-circular orbits.
Classical physics removed from the fundamental law category also 
the initial conditions for the electromagnetic field and 
all other forms of matter and energy (responsible for almost all the complexity we observe), 
leaving the fundamental laws quite simple.
A TOE with a landscape and inflation reclassifies many of the remaining ``laws'' as initial conditions,
since they can differ from one post-inflationary region to another, but since 
inflation generically makes each such region infinite, it can 
fool us into misinterpreting these environmental properties as fundamental laws.
Finally, if the Level IV multiverse of all mathematical structures \cite{toe} exists, 
then even the ``theory of everything'' equations that physicists are seeking are merely local bylaws
in Rees' terminology\cite{ReesHabitat}, differing across the ensemble.
}
\end{figure}

Around when I started high school in 1982, it was discovered that 
a process known as {\it inflation}, 
involving a nearly exponential stretching of space, 
could solve these and other problems  
in one fell swoop \cite{Guth81,Starobinsky1980,Linde82,AlbrechtSteinhardt82,Linde83},
and it soon emerged as the most popular theory for what happened very early on.
Inflation is simple and elegant, requiring merely the existence of some form 
of matter that stubbornly refuses to have its density diluted as space expands
(see \cite{LindeBook,DodelKinneyKolb97,LythRiotto99,LiddleLythBook,Peiris03,Kinney03,LiddleSmith03,Wands03}
for reviews).
The cosmic density fluctuations are explained as the quantum fluctuations required by the Heisenberg uncertainty principle,
magnified by the stretching of space and amplified by gravity.

Previous breakthroughs in theoretical physics like relativity theory and quantum mechanics not only
solved old problems, but also transformed and deepened our understanding of the nature of physical reality.
Inflation did the same.
First of all, it soon became clear that inflation is generically eternal
\cite{LindeBook,Vilenkin83,Starobinsky84,Starobinsky86,Goncharov86,SalopekBond91,LindeLindeMezhlumian94},
so that even though inflation has ended in the part of space that we inhabit (\fig{SphereFig}), it still continues elsewhere and
will ultimately produce an infinite number of other post-inflationary volumes as large as ours, forming a cosmic fractal of sorts.

Moreover, as illustrated in \fig{BoundaryFig}, independent progress in theoretical physics has gradually shifted the borderline between
``laws of physics'' and ``initial conditions'' at the expense of the former.
For example, a common feature of much string theory related model building is that 
there is a ``landscape'' of solutions, corresponding to spacetime configurations involving 
different dimensionality, different types of fundamental particles and different values for certain physical ``constants''.
As an example, Table 1 lists the 30 parameters specifying the standard models of particle physics and cosmology, 
some or all of which may vary across the landscape. 
Eternal inflation transforms such potentiality into reality, actually creating regions of space realizing each of
these possibilities. However, each such region where inflation has ended is generically infinite in size, making 
it impossible for any inhabitants to travel to other regions where these apparent laws of physics are different.

\begin{table}
\noindent 
{\footnotesize
Table~1: The 19 parameters of the $SU(3)\times SU(2)\times U(1)$
standard model of particle physics, compiled from \cite{PDG}, followed by 11 cosmological parameters
as compiled in \cite{inflation}.
Massive neutrinos require additional parameters. Planck units are used, and 
$\mu^2$ and $\lambda$ are defined so that the Higgs potential is $V(\Phi)=\mu^2|\Phi|^2+\lambda|\Phi|^4$.	
\begin{center}
{\tiny
\begin{tabular}{|l|ll|}
\hline
Parameter		&Meaning				 &Measured value\\
\hline
$g		$	&Weak coupling constant			 &$0.6425$\\  
$\theta_W	$	&Weinberg angle				 &$0.4908$\\  
$g_s		$	&Strong coupling constant		 &$\approx 1.2$\\ 
\cline{2-3}
$\mu^2		$	&Quadratic Higgs coefficient		 &$\sim -10^{-33}$\\ 
$\lambda	$	&Quartic Higgs coefficient		 &$\sim 1$?\\ 
\cline{2-3}
$G_e            $	&Electron Yukawa coupling		 &$2.94\times 10^{-6}$\\
$G_\mu          $	&Muon Yukawa coupling			 &$0.000607$\\
$G_\tau         $	&Tauon Yukawa coupling			 &$0.0102156233$\\
\cline{2-3}
$m_u            $	&Up quark Yukawa coupling		 &$0.000016\pm 0.000007$\\
$m_d            $	&Down quark Yukawa coupling		 &$0.00003\pm 0.00002$\\
$m_c            $	&Charm quark Yukawa coupling		 &$0.0072\pm 0.0006$\\ 
$m_s            $	&Strange quark Yukawa coupling		 &$0.0006\pm 0.0002$\\
$m_t            $	&Top quark Yukawa coupling		 &$1.002\pm 0.029$\\ 
$m_b            $	&Bottom quark Yukawa coupling		 &$0.026\pm 0.003$\\
\cline{2-3}
$\sin\theta_{12}$	&Quark CKM matrix angle 		 &$0.2243\pm 0.0016$\\
$\sin\theta_{23}$	&Quark CKM matrix angle 		 &$0.0413\pm 0.0015$\\
$\sin\theta_{13}$	&Quark CKM matrix angle 		 &$0.0037\pm 0.0005$\\
$\delta_{13}	$	&Quark CKM matrix phase 		 &$1.05\pm 0.24$\\
\cline{2-3}
$\thetaqcd	$	&CP-violating QCD vacuum phase 		 &$<10^{-9}$\\
\hline   
$\xib		$	&Baryon mass per photon $\rhob/\ng$				&$(0.49\pm 0.03)\times 10^{-28}$\\
$\xic		$	&CDM mass per photon $\rhoc/\ng$				&$(2.7\pm 0.2)\times 10^{-28}$\\
$\xin		$	&Neutrino mass per photon $\rhon/\ng={3\over 11}\sum m_{\nu_i}$	&$<0.9\times 10^{-28}$\\ 
\cline{2-3}
$\Otot       	$	&Spatial curvature						&$1.01\pm 0.02$\\ 
$\rhol		$	&Dark energy density						&$(9.3\pm 2.5)\tento{-124}$\\ 
$w		$	&Dark energy equation of state					&$-1\pm 0.1$\\ 
$\dH       	$	&Scalar fluctuation amplitude $\delta_H$ on horizon		&$(2.0\pm 0.2)\times 10^{-5}$\\ 
$\ns	        $	&Scalar spectral index					        &$0.98\pm 0.02$\\ 
$\alpha 	$       &Running of spectral index $d\ns/d\ln k$		        &$|\alpha|\simlt 0.01$\\ 
$r           	$	&Tensor-to-scalar ratio	$(\dHt/\dH)^2$      	        	&$\simlt 0.36$\\ 
$\nt           	$ 	&Tensor spectral index						&Unconstrained\\
\hline
\end{tabular}
} 
\end{center}     
} 
\end{table}

If we define parallel universes as regions that are for all practical purposes 
disconnected (outside of causal contact for much longer than the lifetime of any observers), then
these post-inflationary regions are an example thereof. They are labeled as ``Level II" in 
Figure~\ZoomFig,
which is my attempt from \cite{toe} to classify various types of parallel universes that have been 
discussed in the literature. Since each such Level 2 parallel universe is infinite in size, 
it consists of infinitely many spheres as in \fig{SphereFig} --- this Level I multiverse
is much less diverse, with the only difference between the spheres being 
the initial matter distribution (the initial conditions in the limited sense of 
classical physics --- see \fig{BoundaryFig}).
Since the inflationary fluctuations are of quantum origin, inflation also populates the Level III multiverse
(if quantum mechanics is applicable to this multiverse as a whole). We will return to Level IV below.

So how did it all begin? Although inflation gives a beautifully unified answer to the conundra of the 1970's,
we have seen that it still only refines the initial query further, leaving us with a number of questions:
\begin{enumerate}
\item How can we test whether inflation really happened?
\item What is the physics underlying inflation? (What is this hard-to-dilute substance?)
\item Why has inflation recently restarted? (What is the dark energy currently accelerating our universe?)
\item How did inflation begin? Or did it?
\end{enumerate}
Thus addressing the question of how it all began is highly relevant also to other key questions
in physics, such as the quest for the correct theory at the highest energies.
Probing inflation might offer our best hope to test string theory
and other quantum gravity candidates, 
since the early Universe is an unmanned physics experiment probing 
energy scales vastly exceeding those accessible in laboratories.

\section{Observational tests}

This is an exciting time to tackle these question, because 
cosmological observations are finally becoming sensitive enough to help answer them.
All inflation models solve the above-mentioned pre-1980 problems (the flatness problem, the horizon problem, \etc).
In addition, as illustrated in \fig{1DconditionedFig} and elaborated in \cite{inflation}, 
inflation may explain the values of as many as eight observable cosmological parameters
(the last eight in Table 1), including those associated with dark energy.
In the last few years, an avalanche of new cosmological data has revolutionized our ability
to measure these parameters using tools such as the cosmic microwave background (CMB),
galaxy clustering, gravitational lensing, the Lyman alpha forest, cluster abundances and type Ia supernovae
\cite{Spergel03,sdsspars,sdsslyaf,2dfpars,sdssbump}, and I have worked hard to help 
carry this out in practice.
In one suite of papers, I developed methods for analyzing cosmological data sets using
information theory (\eg, \cite{karhunen,galfisher,galpower,mapmaking,cl,polarization,strategy}
and applied them to various CMB experiments and galaxy redshift surveys (\eg, \cite{cobepow,saskmap,2df,sdsspower}),
often in collaboration with the experimentalists/observers who had gathered the data.
Another series of papers tackled various ``dirty laundry'' issues such
as microwave foregrounds and mass-to-light bias (\eg, \cite{mapforegs,wiener,foregrounds,foregpars,bias,r}).
Other papers developed and applied techniques for clarifying the big picture in cosmology:
comparing and combining diverse cosmological probes, cross-checking for
consistency and constraining cosmological models and their free parameters 
(\eg, \cite{9par,10par,boompa,concordance,consistent,sdsspars}).

\clearpage
\begin{figure}[tbp]
\vskip-1.5cm
\centerline{{\vbox{\hglue-0.5cm\epsfxsize=19.4cm\epsfbox{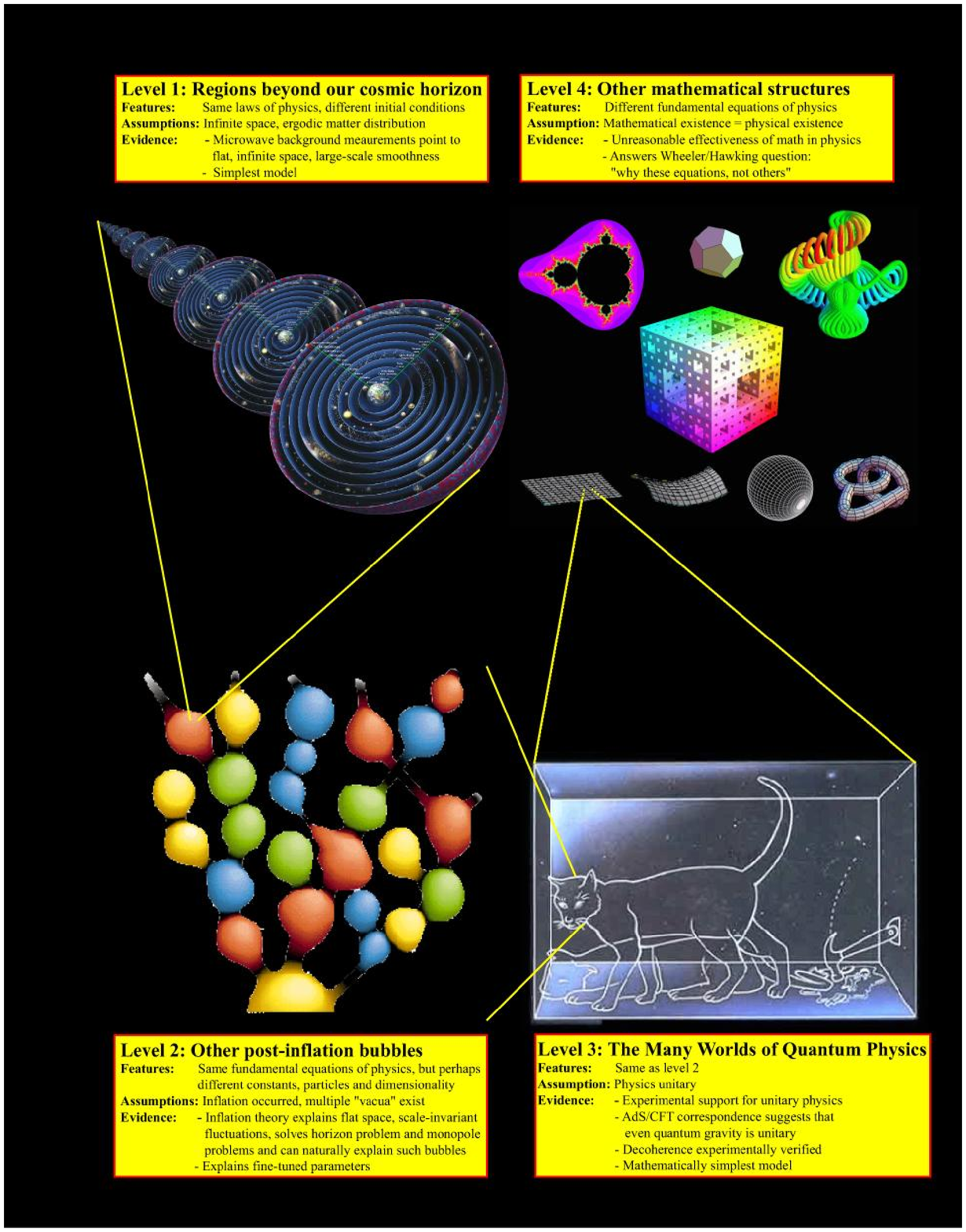}}}}
\end{figure}
\setcounter{figure}{4}
\clearpage

\begin{figure} 
\centerline{\epsfxsize=\figsize\epsffile{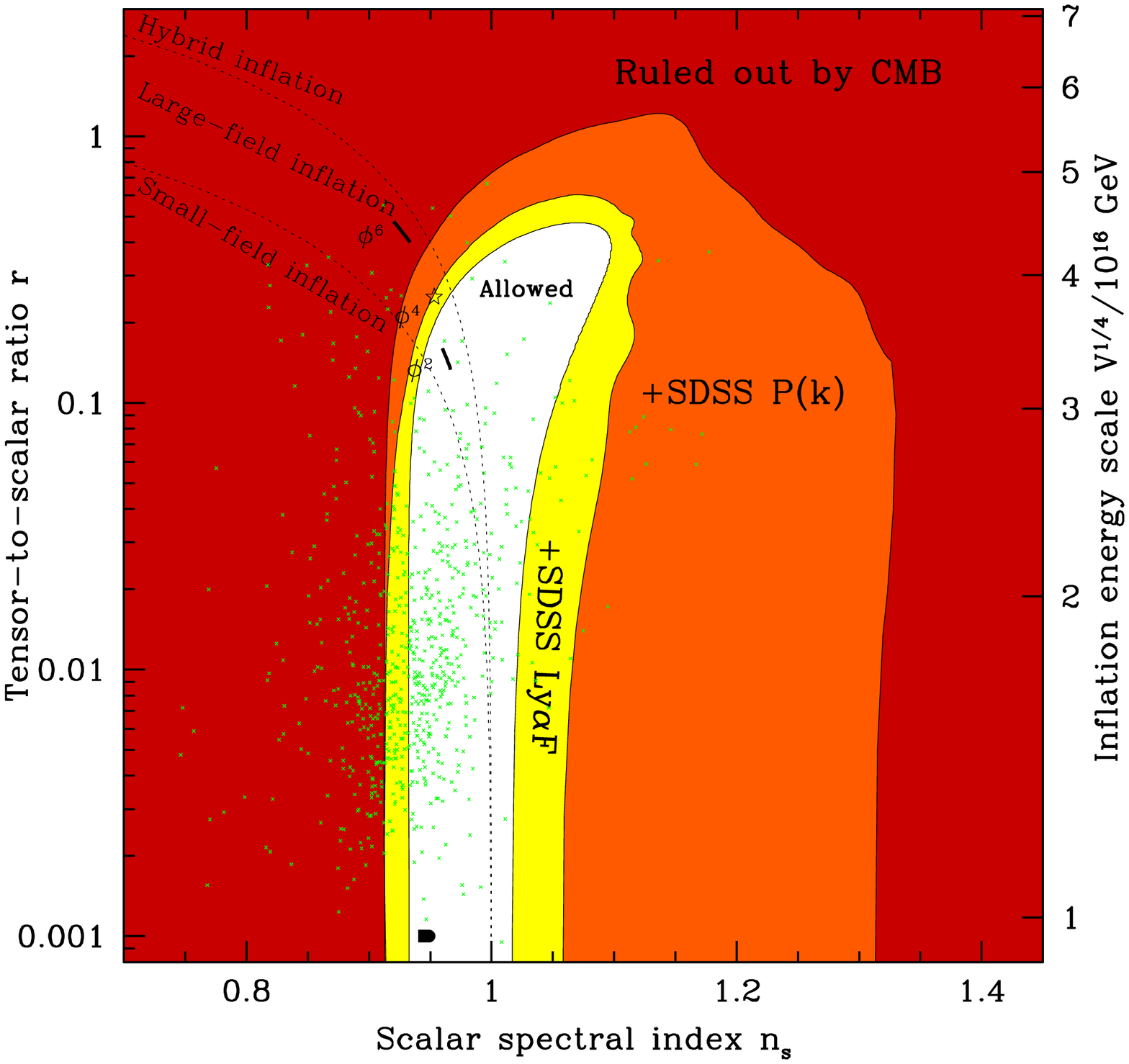}}
\caption[1]{\label{nsrFig}\footnotesize%
Constraints and predictions in the $(\ns,r)$-plane show that
observations are now finally starting to bump up against inflation theory in
an interesting way.
The nested shaded regions are ruled out at 95\% confidence from WMAP CMB observations alone \cite{Spergel03},
when adding SDSS galaxy clustering information \cite{sdsspower,sdsspars} and
when also adding SDSS Lyman $\alpha$ Forest information \cite{sdsslyaf}.
The green/grey points are the predictions from one of my simulations as described in \cite{inflation},
many producing a gravitational wave amplitude $r$ large enough to be detectable 
by combining SDSS with the Planck CMB satellite in 2010 and measuring both $\ns$ and $r$ to an accuracy around 0.01
\cite{parameters2}.
}
\end{figure}

A robust prediction common to essentially all inflation models is that we should measure 
negligible curvature $\Otot=1\pm 10^{-5}$, strikingly confirmed by recent precision measurements such as
$\Otot=1.01\pm 0.02$ \cite{sdsspars} and $\Otot=1.01\pm 0.01$ \cite{sdssbump}. 
Most models also predict approximately scale-invariant seed fluctuations ($\ns\approx 1$), in 
good agreement with the recent measurement $\ns=0.98\pm 0.03$ \cite{sdsspars} (\fig{nsrFig}).
However, data are now getting sensitive enough to look for small departures from
``vanilla'' (scale-invariant, scalar, adiabatic and Gaussian) fluctuations, at least one of which is
expected for essentially all published inflation models, so it is important and timely to work out the detailed predictions
of competing models.

\begin{figure} 
\vskip-4.1cm
\centerline{\epsfxsize=\figsize\epsffile{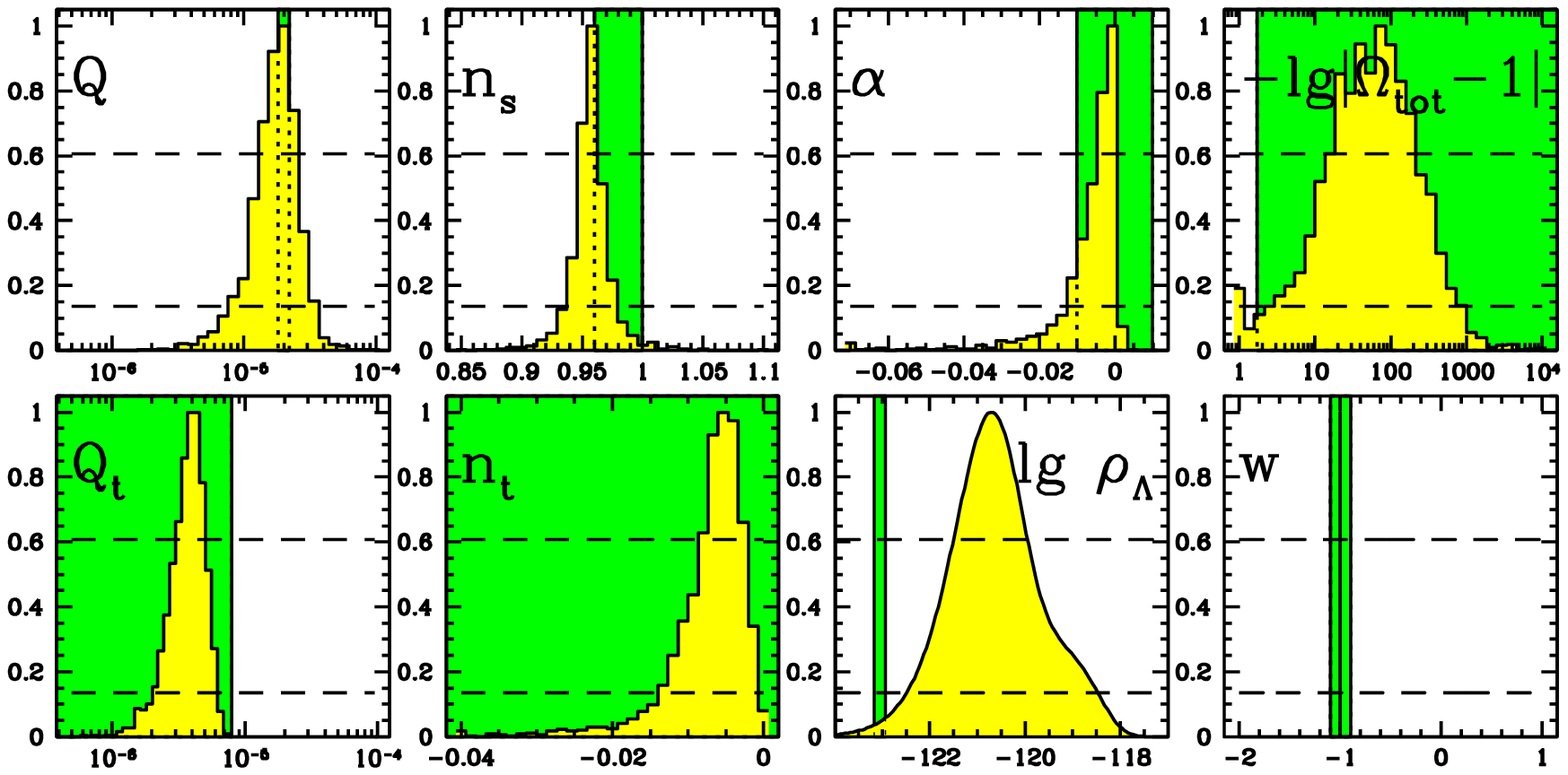}}
\vskip-1cm
\caption[1]{\label{1DconditionedFig}\footnotesize%
Yellow/light grey cosmological parameter distributions show inflationary predictions 
for one of my examples from \cite{inflation}.
Green/dark grey regions show observational constraints ($1\sigma$) \cite{sdsspars,sdsslyaf}.
$\rhol$ is in Planck units.
}
\end{figure}

However, this theoretical calculation is proving surprisingly difficult!
The reason is that most models predict a complicated spacetime with infinitely many observers 
\cite{LindeBook,Vilenkin83,Starobinsky84,Starobinsky86,Goncharov86,SalopekBond91,LindeLindeMezhlumian94},
some of which measure different parameter values from others.
The first 25 parameters in Table 1 will at least be constant within each Level II universe,
but the last 5 can vary even between Level I universes, depending not only on which potential energy 
minimum the so-called inflaton field(s) rolled down into, but also on the path by which it got there.
As discussed in \cite{inflation}, this means that generic inflation models (all models except ones 
that are perfectly symmetric around a single unique minimum) will predict not definite parameter values, 
but merely a probability distribution as in \fig{1DconditionedFig}.
Moreover, as elaborated in \cite{inflation}, computing this probability distribution for what an observer should expect to measure
(and hence making inflation testable)
requires solving technical problems directly involving the more philosophical-sounding aspects of inflation:
\begin{enumerate}
\item The answer depends on the definition of ``observer'' (whether we compute the parameter distribution seen
from a random point, a random proton, a random galaxy, a random planet, \etc).
\item The answer depends on the order(!) in which the infinitely many observers are counted.
\item The answer may depend on pre-inflationary initial conditions.
\end{enumerate}
In spite of early difficulties, the daunting problem of how to predict 
probabilities was successfully overcome in both
classical statistical mechanics and quantum mechanics, making me hopeful that inflation
will follow suit.
Building on \cite{inflation}, I am currently tacking these problems on several fronts, together with colleagues, as well as pursuing more hands-on calculations.

\clearpage
\section{Outlook}

In the endevour to understand where everything comes from, 
two partial answers have in my opinion been found:
\begin{itemize}
\item Q: Where does the observed {\it matter} come from?\\
      A: Inflation can produce it all from almost nothing.
\item Q: Where does the observed {\it complexity} come from?\\
      A: Parallel universes
can produce it all from almost nothing, with the fundamental laws being simple and almost all
the complexity existing only in the mind of the beholder, since the individual parallel universes
require vastly more information to describe than the multiverse as a whole \cite{nihilo}.
\end{itemize}

In conclusion, the age-old question ``How did it all begin'' 
has been dramatically refined in recent years, transformed into a quest to 
understand cosmological inflation and physics at the highest energies.
In this quest, parallel experimental and theoretical progress has fruitfully connected mainstream 
empirical work to fundamental theoretical research that in turn has
profound philosophical implications regarding 
our cosmic origin, our cosmic future, fundamental/environmental laws and parallel universes.
In other words, looking ahead, if has never been more interesting than now to ask how it all began.

\bigskip
{\bf Acknowledgements:}
I wish to thank Anthony Aguirre, Ang{\'e}lica de Oliveira-Costa, Martin Rees and Frank Wilczek for inspiring 
discussions.
This work was supported by NASA grant NAG5-11099,
NSF CAREER grant AST-0134999, and fellowships from the David and Lucile
Packard Foundation and the Research Corporation.

\end{document}